\documentstyle[preprint,epsfig,aps,floats]{revtex}

\begin{document}

\tighten
%% astro-ph/0007329
\preprint{\font\fortssbx=cmssbx10 scaled \magstep2
\hbox to \hsize{
\hbox{\fortssbx University of Wisconsin - Madison}
\hfill$\vcenter{\tighten
                \hbox{\bf MADPH-00-1176}
                \hbox{July 2000}}$}}

\title{\vspace{.5in}
$10^{20}$\,eV cosmic-ray and particle physics with kilometer-scale
neutrino telelscopes}

\author{J. Alvarez-Mu\~niz and F. Halzen}

\address{Univ. of Wisconsin, Dept. of Physics, 1150 University Avenue, Madison,
Wisconsin 53706, USA.}

\maketitle

\begin{abstract}
We show that a kilometer-scale neutrino observatory, though optimized for TeV  
to PeV energy, is sensitive to the neutrinos associated 
with super-EeV sources.  
These include super-heavy relics, neutrinos associated with the Greisen  
cutoff, and topological defects which are remnant cosmic structures associated  
with
phase transitions in grand unified gauge theories. It is a misconception that  
new instruments optimized to EeV energy are required to do this important  
science, although this is not their primary goal. Because 
kilometer-scale neutrino telescopes can reject atmospheric  
backgrounds by establishing the very high energy of the signal events, they  
have sensitivity over the full solid angle, including the horizon 
where most of  
the signal is concentrated. This is important because up-going  
neutrino-induced muons, routinely considered in previous calculations, are  
absorbed by the Earth.

\end{abstract}

\vskip 0.5cm

PACS number(s): 95.55.Vj, 96.40.Tv, 98.70.Sa, 98.80.Cq  

%{\bf Keywords:}

\section{Introduction}

It has been realized for some time that topological defects are unlikely to be  
the origin of the structure in the present Universe \cite{TDstructure}.
Therefore the observation
of their decay products, in the form of cosmic rays or high energy neutrinos,  
becomes the most straightforward way to search for these remnant structures
from grand unified phase transitions \cite{TD}. 
Such search represents an example of fundamental
particle physics that can only be done with cosmic beams. We here point out
that a kilometer-scale neutrino observatory \cite{physrep}, such as 
IceCube, has excellent
discovery potential for topological defects. The instrument can identify the  
characteristic
signatures in the energy and zenith angle distribution 
of the signal events. It  
is a common misconception that different instruments\cite{auger,owl},  
optimized to EeV
signals, are required to do this important science, although this 
is not their primary motivation. Our conclusions for  
topological defects extend to other physics associated with $10^{20} -  
10^{24}$\,eV energies. 

We will illustrate our claims by demonstrating IceCube sensitivity to:

\begin{itemize}

\item generic topological defects with grand-unified mass scale $M_X$ of order  
$10^{14}-10^{15}$\,GeV and a particle decay spectrum consistent with all present
observational constraints\cite{protheroe,sigl,pillado},
\item superheavy relics, normalized to the Z-burst  
scenario\cite{weiler} where the  
observed ultra high energy cosmic rays (UHECR) of $\sim 10^{20}$\,eV energy  
and above are locally produced by the interaction of superheavy relic
neutrinos with the cosmic neutrino background radiation \cite{gelmini},
\item neutrinos produced by superheavy relics which themselves decay into the  
UHECRs \cite{berez,sarkar}, and
\item the flux of neutrinos produced in the interactions of
UHECR cosmic rays with the microwave background \cite{steckerCMB},
the so called Greisen neutrinos. This flux,  
which originally inspired the concept of a kilometer-scale
neutrino detector, is mostly shown for comparison.

\end{itemize}

The basic reasons for our more optimistic conclusions about the sensitivity of a
detector such as IceCube are simple. Unlike first-generation neutrino
telescopes, IceCube can measure energy and can therefore separate very high  
energy  
signals from the low energy atmospheric neutrino
background by energy measurement \cite{icecube} (see below). 
The instrument can therefore isolate high
energy events over $4\pi$ solid angle, and not just in the hemisphere where
the neutrinos are identified by their penetration of the Earth. This is of
primary importance here because neutrinos from topological defects have
energies high enough so that they are efficiently absorbed by the 
Earth \cite{gandhi}. The
signal from above and near the horizon typically dominates the up-going
neutrino fluxes by an order of magnitude. We will show that the zenith angle
distribution of neutrinos associated with topological defects form a
characteristic signature for their extremely high energy origin.

\section{Neutrino events}

We calculate the neutrino event rates by convoluting the $\nu_\mu+\bar\nu_\mu$  
flux from the different sources considered in this paper, with the probability 
of detecting a muon produced in a  
muon-neutrino interaction in the Earth or atmosphere:

\begin{equation}
N_{\rm events}=2\pi~A_{\rm eff}~T~\int\int~{dN_\nu\over dE_\nu}(E_\nu)
P_{\nu\rightarrow\mu}(E_\nu,E_\mu {\rm (thresh)},\cos\theta_{\rm zenith})
~dE_\nu~d\cos\theta_{\rm zenith}
\end{equation}
where $T$ is the observation time and $\theta_{\rm zenith}$ the zenith 
angle. We assume an effective telescope area of  
$A_{\rm eff}=1~{\rm km^2}$,
a conservative assumption for the
very high energy neutrinos considered here.
It is important to notice that the probability ($P_{\nu\rightarrow\mu}$) 
of detecting a muon
with energy above a certain energy threshold $E_\mu$(threshold),
produced in a muon-neutrino interaction,
depends on the angle of incidence of the neutrinos. This is because the  
distance traveled by a muon cannot exceed the column density of matter available
for neutrino interaction, a condition not satisfied by very high energy  
neutrinos produced in the atmosphere. They are absorbed by the Earth and only  
produce neutrinos in the ice above, or in the atmosphere or Earth near the  
horizon. The event rates in which the muon arrives at the detector 
with an energy above $E_\mu$(threshold)=1 PeV, where the atmospheric neutrino 
background is negligible, are shown in Table I. 

Fig.\,1 shows the
$\nu_\mu+\bar\nu_\mu$ fluxes used in the calculations. We first calculate  
the event rates corresponding to the largest flux from topological  
defects \cite{protheroe} allowed by constraints imposed 
by the measured diffuse  
$\gamma$-ray
background in the vicinity of 100 MeV.  
The corresponding proton flux has been  
normalized to the observed cosmic ray spectrum
at $3\times 10^{20}$\,eV; see Fig.\,2 of reference \cite{protheroe}. 
Models with p$<$1 appear to
be ruled out \cite{sigl} and hence they are not considered in the
calculation. As an  
example of neutrino production by superheavy relic particles, we consider the  
model of  
Gelmini and Kusenko \cite{gelmini}. In Figs.\,2 and 3 we show the event rates  
as a function of neutrino energy. We assume a muon energy threshold
of 1 PeV. 
We also show in both plots the 
event rate due to the Waxman and Bahcall bound \cite{wblimit}. 
This bound represents the maximal
flux from astrophysical, optically thin sources, in which neutrinos are 
produced in p-p or p-$\gamma$ collisions. The atmospheric
neutrino events are not shown since they are negligible above the 
muon energy threshold we are using. The area under the curves 
in both Figs. is equal to the number of events for each source.   
In Fig.\,4 we plot the
event rates in which the produced muon arrives at the detector
with an energy greater than $E_\mu$(threshold). In  
Fig.\,5 we finally present the angular distribution of the neutrino 
events for  
the different very high energy neutrino sources. The characteristic shape of  
the distribution reflects the opacity of the Earth to high energy neutrinos,  
typically  
above $\sim$100 TeV. The limited column density of matter in the atmosphere  
essentially reduces the rate of downgoing neutrinos to interactions in the  
1.5\,km of ice above the detector. The events are therefore concentrated  
near the horizontal direction corresponding to
zenith angles close to $90^{\rm o}$. The neutrinos predicted by the model of  
Gelmini and Kusenko are so energetic that they are even absorbed 
in the horizontal direction as can be seen in Fig.\,5.

\vskip 0.3cm
\begin{center}
\begin{tabular}{|c|c|c|} \hline
Model & $N_{\nu_\mu+\bar\nu_\mu}$ (downgoing) & $N_{\nu_\mu+\bar\nu_\mu}$
(upgoing)  \\ \hline
TD, $M_X=10^{14}$ GeV, $Q_0=6.31\times 10^{-35}$, p=1~ & 11 & 1 \\
TD, $M_X=10^{14}$ GeV, $Q_0=6.31\times 10^{-35}$, p=2~ & 3 & 0.3 \\ \hline
TD, $M_X=10^{15}$ GeV, $Q_0=1.58\times 10^{-34}$, p=1~ & 9  & 1 \\
TD, $M_X=10^{15}$ GeV, $Q_0=1.12\times 10^{-34}$, p=2~ & 2 & 0.2 \\ \hline
Superheavy Relics Gelmini {\it et al.} \cite{gelmini} & 30 & $1.5\times 10^{-7}$ \\  
\hline
Superheavy Relics Berezinsky {\it et al.} \cite{berez} & 2 & 0.2 \\ \hline
Superheavy Relics Birkel {\it et al.} \cite{sarkar} & 1.5 & 0.3 \\ \hline
p-$\gamma_{\rm CMB}$ $(z_{\rm max}=2.2)$ \cite{steckerCMB} & 1.5 & $1.2\times 10^{-2}$ \\ \hline
W-B limit $2\times 10^{-8}~E^{-2}~{\rm (cm^2~s~sr~GeV)^{-1}}$ & 8.5 & 2 \\ \hline \hline
%AGN jets ($\rm yr^{-1}~km^{-2}$) & 1370 & 145 \\ \hline
Atmospheric background & $2.4\times 10^{-2}$ & $1.3\times 10^{-2}$ \\ \hline
\end{tabular}
\end{center}
{\bf Table I:} Neutrino event rates (per year per ${\rm km^2}$ in
$2\pi$ sr) in which the produced muon arrives
at the detector with an energy
above $E_\mu$(threshold)=1 PeV. Different neutrino sources have been
considered.
The topological defect models (TD) correspond to highest injection rates
$Q_0~({\rm ergs~cm^{-3}~s^{-1}})$
allowed in Fig.\,2 of \cite{protheroe}.
Also shown is the number of events from p-$\gamma_{\rm CMB}$ interactions
in which protons are propagated up to a maximum redshift $z_{\rm max}=2.2$ 
\cite{steckerCMB} and the number of neutrinos from the Waxman and Bahcall
limit on the diffuse flux from optically thin sources \cite{wblimit}.
The number of atmospheric background events above 1 PeV is also shown.
The second column corresponds to
downward going neutrinos (in $2\pi$ sr).
The third column gives the number of upward going
events (in $2\pi$ sr). We have taken
into account absorption in the Earth according to
reference \cite{gandhi}. IceCube will detect the sum of the event rates given  
in the last two columns.
\vskip 0.5cm

Energy measurement is critical for achieving the sensitivity of the detector  
claimed. For muons, the energy resolution
of IceCube is anticipated to be $25\%$ in the logarithm of the energy,
possibly better. The detector is able to determine energy to better than
an order of magnitude, sufficient for the separation of EeV signals from
atmospheric neutrinos with energies below 100 TeV. Notice that
one should also be able to identify electromagnetic showers initiated by
electron and tau-neutrinos. Their energy measurement is linear and
expected to be better than $20\%$. Such EeV events will be gold-plated,
unfortunately their fluxes are expected to be even lower. For instance
for the first TD model in Table I (p=1, $M_X=10^{14}$ GeV and 
$Q_0=6.31 \times 10^{-35}~{\rm ergs~cm^{-3}~s^{-1}}$), we expect 
$\sim 1$ contained shower 
per year per ${\rm km^2}$ above 1 PeV initiated in charged 
current interactions of $\nu_e+\bar\nu_e$. The corresponding number
for the Gelmini and Kusenko flux is $\sim 4~{\rm yr^{-1}~km^{-2}}$.

One should also worry about the fact that a very high energy muon may enter  
the detector with reduced energy because of energy losses. It could become  
indistinguishable from atmospheric background \cite{gaisser}. We have 
accounted for the ionization as well as catastrophic muon energy losses 
which are incorporated in the calculation of the range of the muon. 
In the PeV regime region this energy reduction is roughly one 
order of magnitude, it should be less for  
the higher energies considered here. 

In conclusion, if the fluxes predicted by  
our sample of models for neutrino  
production in the super-EeV region are representative, they should be revealed  
by the IceCube observatory operated over several years.

\section*{Acknowledgements}
We thank J.J. Blanco-Pillado for making available to us his code to obtain
the neutrino fluxes from topological defects and E. Zas for helpful discussions.
This research was supported in part by the US Department of Energy under
grant DE-FG02-95ER40896 and in part by the University of Wisconsin
Research Committee with funds granted by the Wisconsin Alumni Research
Foundation.
J.A. thanks the Department of Physics, University of Wisconsin, Madison and the
Fundaci\'on Caixa Galicia for financial support.

\newpage
\begin{figure}[hbtp]
\begin{center}
\mbox{\epsfig{file=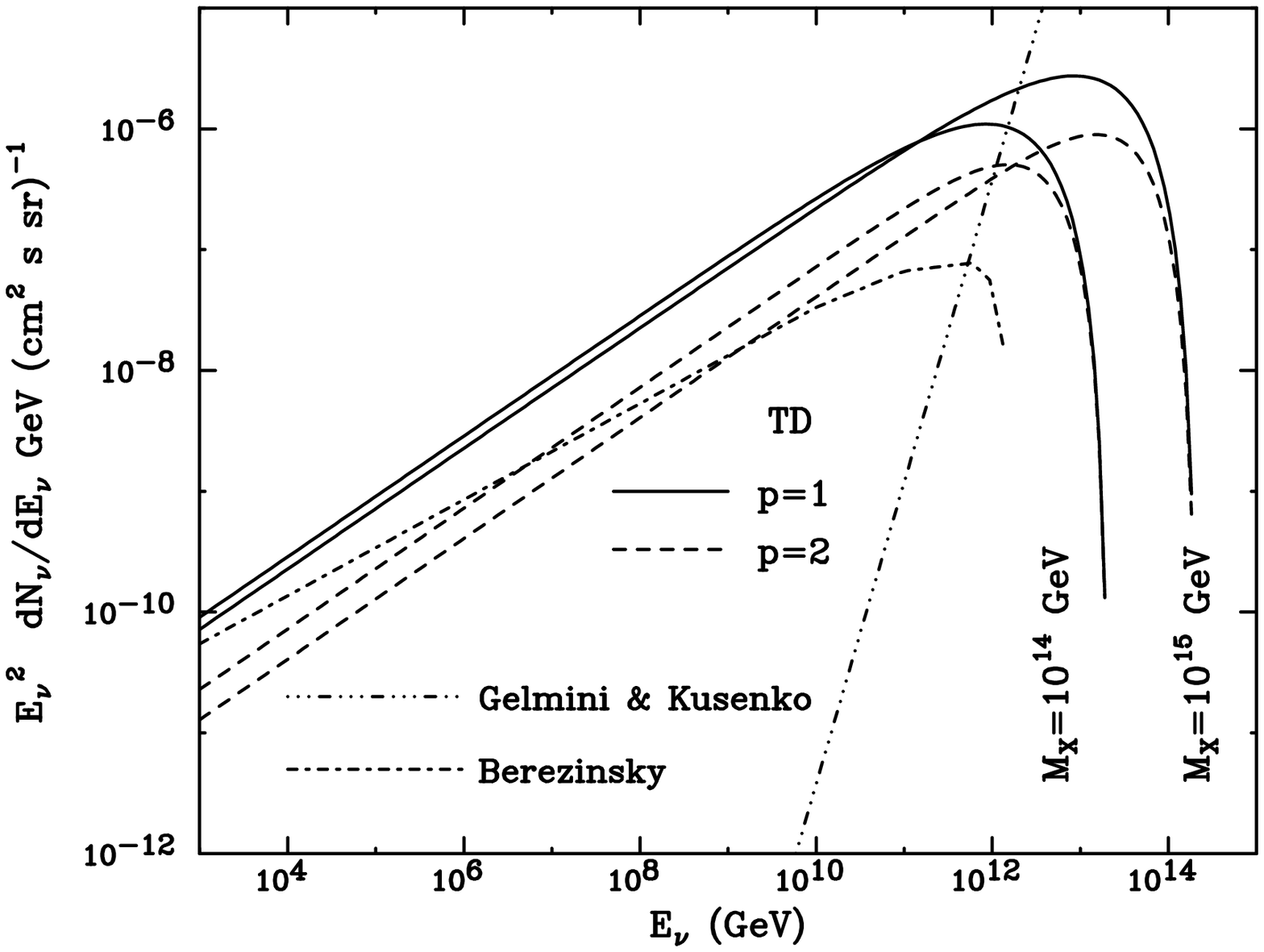,height=10cm}}
\end{center}
\caption{Maximal predictions of
$\nu_\mu+\bar\nu_\mu$ fluxes from topological defect models
by Protheroe and Stanev (p=1,2). Also
shown is the $\nu_\mu+\bar\nu_\mu$ from superheavy relic particles
by Gelmini and Kusenko and the flux by Berezinsky {\it et al.}.}
\end{figure}

\newpage
\begin{figure}[hbtp]
\begin{center}
\mbox{\epsfig{file=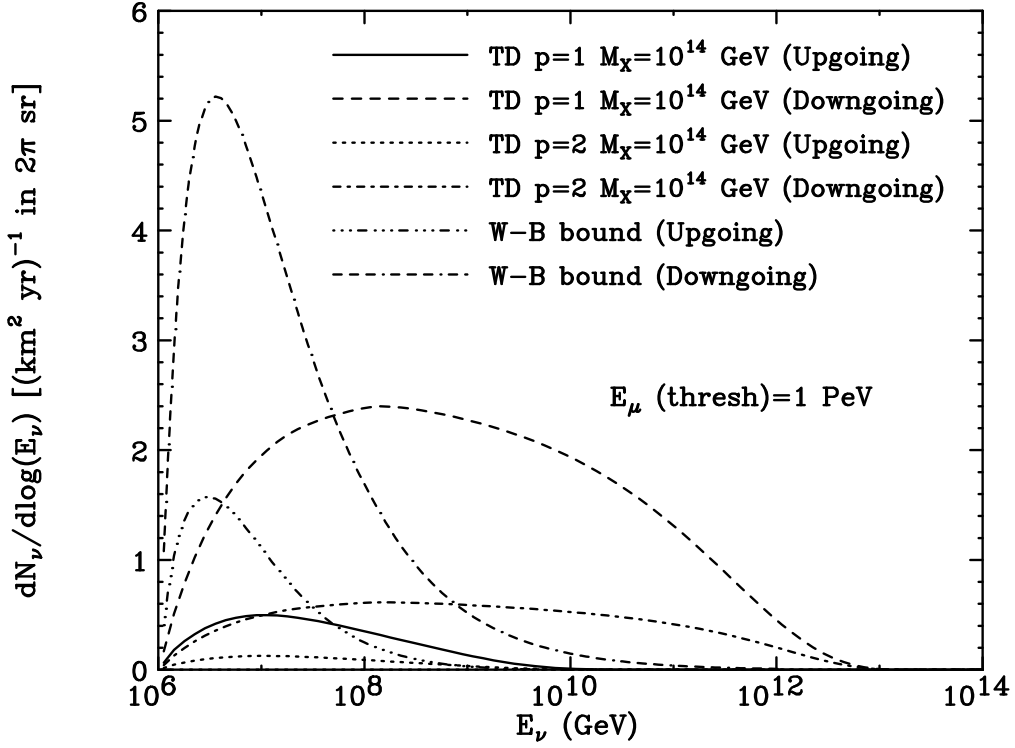,height=10cm}}
\end{center}
\caption{Differential $\nu_\mu+\bar\nu_\mu$ event rates in IceCube from
the topological defect fluxes in Fig.1.
The muon threshold is $E_\mu$(threshold)=1 PeV. 
We have separated the contribution from 
upgoing and downgoing events to stress the different behavior with 
energy. The event rate expected from the Waxman and Bahcall bound (see text)
is also shown for illustrative purposes. The rate due to atmospheric
neutrinos is negligible (see Table I) and hence it is not plotted.}
\end{figure}

\newpage
\begin{figure}[hbtp]
\begin{center}
\mbox{\epsfig{file=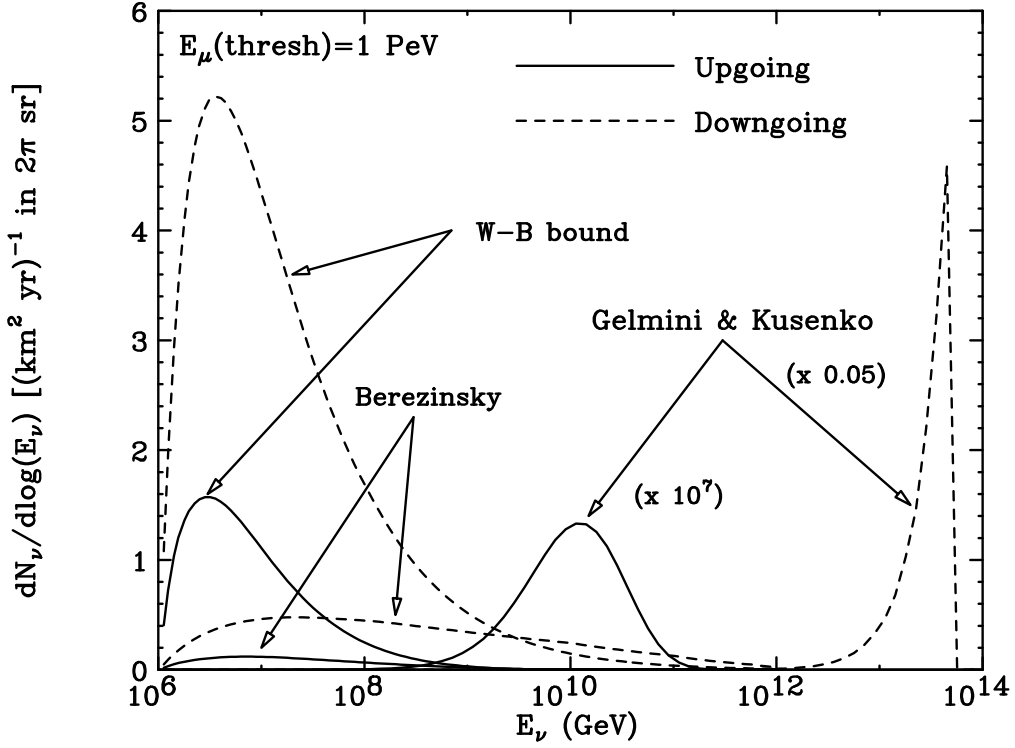,height=10cm}}
\end{center}
\caption{Differential $\nu_\mu+\bar\nu_\mu$ event rates in IceCube
from super-heavy relic particles. We have separated the contribution 
from upgoing and downgoing events to stress the different behavior with
energy.   
The muon threshold is $E_\mu$(threshold)=1 PeV. 
The event rate due to atmospheric 
neutrinos as well as the one expected from the Waxman and 
Bahcall bound (see text) is shown for illustrative purposes.
The rate due to atmospheric
neutrinos is negligible (see Table I) and hence it is not plotted.}
\end{figure}

\newpage
\begin{figure}[hbtp]
\begin{center}
\mbox{\epsfig{file=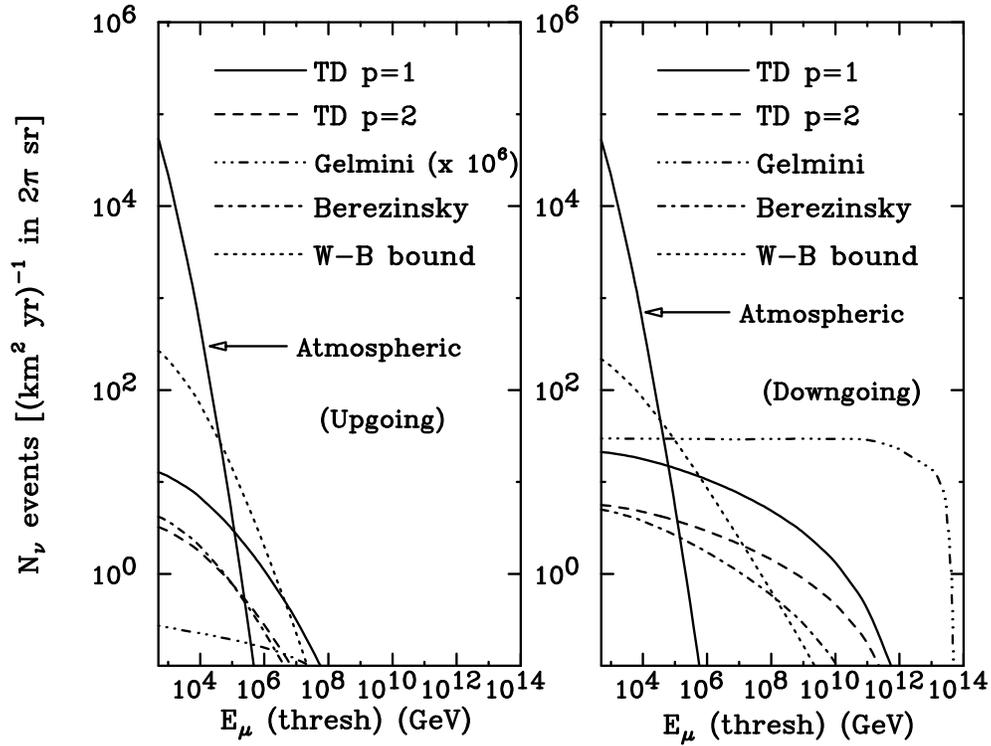,height=10cm}}
\end{center}
\caption{$\nu_\mu+\bar\nu_\mu$ event rates in IceCube from
the fluxes in Fig.1. The plot shows the number of events in which 
the produced muon arrives at the detector with an energy above $E_\mu$(thresh). 
Atmospheric neutrino events 
and the event rate expected from the Waxman and Bahcall upper bound
(see text) are also plotted. The topological defect (TD) models shown
(p=1 and p=2) correspond to $M_X=10^{14}$ GeV.
Upgoing and downgoing events are shown separately.} 
\end{figure}

\newpage
\begin{figure}[hbtp]
\begin{center}
\mbox{\epsfig{file=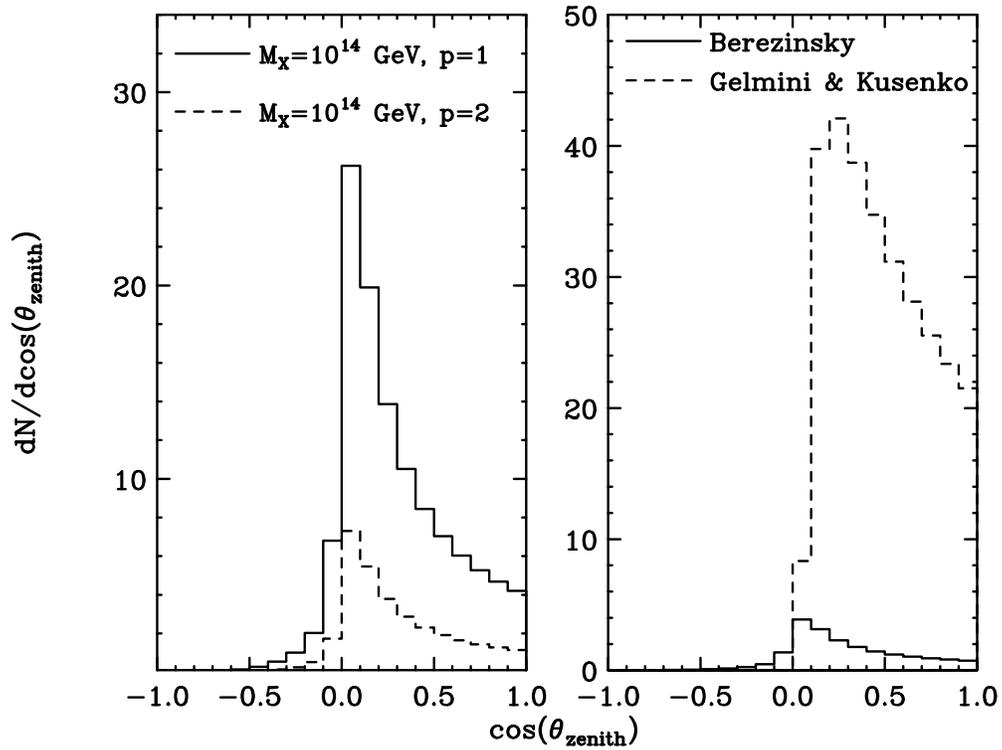,height=10cm}}
\end{center}
\caption{Zenith angle distribution of the  
$\nu_\mu+\bar\nu_\mu$ events in IceCube in which the produced muon arrives
at the detector with energy above 1 PeV.
Left: Topological defect models. Right: Superheavy relics. 
$\cos(\theta_{\rm zenith})=-1$
corresponds to vertical upgoing neutrinos, $\cos(\theta_{\rm zenith})=0$
to horizontal neutrinos and $\cos(\theta_{\rm zenith})=1$ to vertical
downgoing neutrinos. The detector
is located at a depth of 1.8 km in the ice.}
\end{figure}

\end{document}